\documentclass[conf]{new-aiaa}
\usepackage[utf8]{inputenc}

\usepackage{graphicx}
\usepackage{amsmath}
\usepackage[version=4]{mhchem}
\usepackage{siunitx}
\usepackage{subcaption}
\usepackage{longtable,tabularx}
\newtheorem{remark}{Remark}
\usepackage{soul}

\title{
    Smooth Reference Command Generation and Control for Transition Flight of VTOL Aircraft Using Time-Varying Optimization
}

\author{
Jinrae Kim\footnote{Postdoctoral Research Associate, Department of Mechanical Science and Engineering,
The Grainger College of Engineering,
University of Illinois Urbana-Champaign.},
John L. Bullock\footnote{Doctoral Student, Department of Electrical and Computer Engineering,
The Grainger College of Engineering,
University of Illinois Urbana-Champaign.},
Sheng Cheng\footnote{Postdoctoral Research Associate, Department of Mechanical Science and Engineering,
The Grainger College of Engineering,
University of Illinois Urbana-Champaign.},
and Naira Hovakimyan\footnote{Professor, Department of Mechanical Science and Engineering,
The Grainger College of Engineering,
University of Illinois Urbana-Champaign. AIAA Fellow.}
}

\begin{document}

\maketitle

\begin{abstract}
Vertical take-off and landing (VTOL) aircraft pose a challenge in generating reference commands during transition flight.
While sparsity between hover and cruise flight modes can be promoted for effective transitions by formulating $\ell_{1}$-norm minimization problems,
solving these problems offline pointwise in time can lead to non-smooth reference commands,
resulting in abrupt transitions.
This study addresses this limitation by proposing a time-varying optimization method that explicitly considers time dependence.
By leveraging a prediction-correction interior-point time-varying optimization framework,
the proposed method solves an ordinary differential equation to update reference commands continuously over time,
enabling smooth reference command generation in real time.
Numerical simulations with a two-dimensional Lift+Cruise vehicle validate the effectiveness of the proposed method,
demonstrating its ability to generate smooth reference commands online.
\end{abstract}

\section{Introduction}
\lettrine{L}{ift}+Cruise vehicles are a class of vertical take-off and landing (VTOL) vehicles that combine the characteristics of fixed-wing aircraft and rotorcraft.
Due to their inherent mixed nature,
Lift+Cruise aircraft are capable of operating in a large flight envelope with agile maneuverability.
In this regard,
algorithms and simulators for autonomy of Lift+Cruise aircraft have been extensively studied~\cite{simmonsFullEnvelopeAeroPropulsiveModel2021,bullockReferenceCommandOptimization2024,leeMPCBasedLongitudinalControl2024,kimVTOLAircraftOptimal2023,guam_v11}.
The flight of Lift+Cruise aircraft can be categorized into three modes:
hover, transition, and cruise~\cite{simmonsFullEnvelopeAeroPropulsiveModel2021}.
Planning and control of hover and cruise modes are relatively straightforward
from the two different origins of rotor-borne and wing-borne flights.
On the contrary,
the transition between hover and cruise can be challenging due to the coupling of two distinctive modes.
This raises an important practical challenge:
how to generate the reference command of actuators from hover to cruise and vice versa~\cite{bullockReferenceCommandOptimization2024}?

Differential flatness can be used to generate reference commands for 2D Lift+Cruise vehicle models~\cite{bullockReferenceCommandOptimization2024}.
Unlike differential flatness of multicopters~\cite{sunComparativeStudyNonlinear2022},
for a given reference position command,
the differential flatness of the 2D Lift+Cruise vehicle model creates ambiguity when determining the total thrusts and the reference pitch command.
To resolve this ambiguity,
an optimization-based method was proposed ~\cite{bullockReferenceCommandOptimization2024}.
In the optimization-based method,
the time horizon is divided into multiple time segments.
Then, it performs $\ell_{1}$-norm minimization pointwise in time and interpolates solutions over time offline.
The $\ell_{1}$-norm minimization encourages sparsity, enabling the transition between hover and cruise modes effectively.
However, sparsity may result in abrupt and non-smooth reference command generation with input saturation during transition.
In practice,
tracking controllers are implemented to track the reference commands generated offline, i.e., open-loop commands.
Non-smooth reference commands often cause a slight tracking error,
and this error can be accumulated due to input saturation.

To overcome the limitations of the existing study~\cite{bullockReferenceCommandOptimization2024},
in this paper
we propose an online reference command generation method.
The proposed method is based on a time-varying optimization formulation in which the time dependency of the optimization problem is explicitly considered.
This formulation allows us to update the optimization variable continuously over time as a solution to an ordinary differential equation (ODE).
Therefore, the proposed method can generate a smooth reference command.
Furthermore, the proposed method can generate reference commands online (in a closed-loop sense) by solving the ODE forward in time.
Numerical simulation results with tracking controllers show that the proposed method can generate smooth reference commands online, avoiding accumulated tracking error due to input saturation compared to an existing open-loop method~\cite{bullockReferenceCommandOptimization2024}.

The rest of this paper is organized as follows:
Section \ref{sec:preliminaries} describes the equations of motion for the Lift+Cruise vehicle model considered in this study and introduces an existing optimization-based reference command generation method. Potential issues of the existing method are addressed,
and the objective of this study is given.
In section \ref{sec:main},
an online reference command generation method is proposed based on time-varying optimization.
In section \ref{sec:sim},
numerical simulation is performed to demonstrate the characteristics and performance of the proposed method.
Section \ref{sec:conclusion} concludes this study.

\section{Preliminaries}
\label{sec:preliminaries}
\subsection{Equations of Motion for 2D Lift+Cruise vehicle}
\begin{figure}
    \centering
    \begin{subfigure}{.5\textwidth}
    \centering
    \includegraphics[width=.8\linewidth]{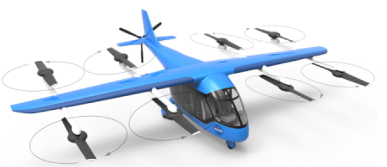}
    \caption{}
    \label{fig:vtol_realistic}
    \end{subfigure}%
    \begin{subfigure}{.5\textwidth}
    \centering
    \includegraphics[width=.8\linewidth]{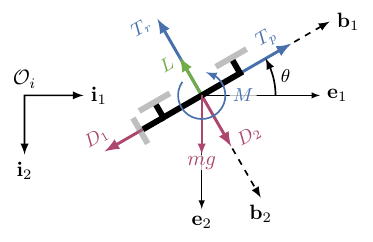}
    \caption{}
    \label{fig:vtol_simplified}
    \end{subfigure}%
    \caption{Illustration of (a) Lift+Cruise (L+C) vehicle and (b) 2D model}
    \label{fig:vtol_illustration}
\end{figure}

Consider a simplified 2D Lift+Cruise vehicle as shown in \autoref{fig:vtol_illustration}.
The equations of motion for the 2D Lift+Cruise vehicle are introduced in~\cite{bullockReferenceCommandOptimization2024}.
In this study, we modify the notation so that the $z$-axis of the inertial frame corresponds to ``Down'' to follow the aerospace convention.
Then, the equations of motion can be written as follows:
\begin{equation}
\label{eq:dyn}
\begin{split}
\ddot{p} &= m^{-1} R(\theta)^{\intercal} F + g e_{2},
\\
\ddot{\theta} &= J^{-1} M,
\end{split}
\end{equation}
where $p := [x, z]^{\intercal}$ is the position in 2D inertial coordinate system with $x$ and $z$ coordinates,
$\theta \in \mathbb{R}$ is the pitch angle,
and $M \in \mathbb{R}$ is the applied torque.
In this study, function arguments such as the independent variable time, $t \in \mathbb{R}$, may be omitted for brevity.
The mass of the vehicle, the moment of inertia, and the gravitational constant are denoted by $m > 0 $, $J > 0$, and $g > 0$, respectively.
The unit vector defining the $z$-axis of the inertial frame is denoted by $e_{2} := [0, 1]^{\intercal}$.
The rotational matrix $R(\theta)$ can be written as
\begin{equation}
    R(\theta) = \begin{bmatrix}
        \cos(\theta) & -\sin(\theta) \\
        \sin(\theta) & \cos(\theta)
    \end{bmatrix}.
\end{equation}
The net force $F$ applied to the vehicle can be written as
\begin{equation}
    F = [T_{p} - D_{1}, -(T_{r} + L - D_{2})]^{\intercal},
\end{equation}
where $T_{p} \geq 0$ and $T_{r} \geq 0$ denote the total thrusts of horizontal pushers and vertical rotors, respectively.
The aerodynamic forces, lift $L \in \mathbb{R}$, horizontal drag $D_{1} \in \mathbb{R}$, and vertical drag $D_{2} \in \mathbb{R}$, are modeled as
\begin{equation}
\begin{split}
    L &= u^{2} C_{L},
    \\
    D_{1} &= u^{2} C_{D_{1}},
    \\
    D_{2} &= v^{2} C_{D_{2}},
\end{split}
\end{equation}
where $[u, v]^{\intercal} \in \mathbb{R}^{2}$ denotes the velocity in body-fixed frame.
Under the assumption of small wind speed,
the velocity can be approximated as $[u, v]^{\intercal} \approx R^{\intercal}(\theta) \dot{p}$.
The aerodynamic coefficients are given as follows:
\begin{equation}
\begin{split}
C_{L} &= C_{L_{0}} + C_{L_{\theta}} \theta + C_{L_{\delta_{e}}} \delta_{e},
\\
C_{D_{1}} &= C_{D_{1, 0}} + C_{D_{1, k}} C_{L}^{2},
\\
C_{D_{2}} &= C_{D_{2, 0}},
\end{split}
\end{equation}
where $\delta_{e} \in [\underline{\delta}_{e}, \overline{\delta}_{e}] \subset \mathbb{R}$ is the deflection angle of elevator
with minimum and maximum values of $\underline{\delta}_{e}$ and $\overline{\delta}_{e}$, respectively.
Then,
the vehicle's state and control input can be combined for state-space representation as $\mathbf{x} := [x, z, \theta, \dot{x}, \dot{z}, \dot{\theta}]^{\intercal} \in \mathbb{R}^{6}$
and $\mathbf{u} = [T_{r}, T_{p}, M, \delta_{e}]^{\intercal} \in \mathbb{R}^{4}$,
respectively,
with the aerodynamic forces being functions of $\theta$, $\delta_{e}$, and $t$.

\subsection{Time-varying optimization}
\label{sec:tv_opt}
Time-varying optimization is a branch of mathematical optimization such that the cost function and constraint functions depend on time~\cite{fazlyabPredictionCorrectionInteriorPointMethod2018}.
A general time-varying optimization formulation can be represented as follows:
\begin{equation}
\begin{split}
    \min_{U} \quad &J(U, t)
    \\
    \text{subject to} \quad g_{i}(U, t) &= 0, \forall i \in \mathcal{I}_{eq},
    \\
    h_{i}(U, t) &\leq 0, \forall i \in \mathcal{I}_{ineq},
\end{split}
\end{equation}
where $J$, $g_{i}$ and $h_{i}$ are the cost function, equality constraints, and inequality constraints.
The objective of the time-varying optimization is to find a time-varying minimizer $U^{\star}(t) \in \arg \min_{U} J(U, t)$, subject to constraints.
In general, it is not tractable to find a time-varying global minimizer.
Therefore, the objective can be relaxed to find a time-varying local minimizer.
For example, in the case of unconstrained time-varying optimization,
the objective is to find $\hat{U}^{\star}(t)$ such that $\nabla_{U} J(U, t) \vert_{U=\hat{U}^{\star}(t)} = 0$.

\subsection{Optimization-based open-loop reference command generation with sparsity}
\label{sec:ol_opt}
Differential flatness can be used to generate reference commands in many aerospace applications~\cite{faesslerThrustMixingSaturation2017,sunComparativeStudyNonlinear2022,mellingerMinimumSnapTrajectory2011a,bryAggressiveFlightFixedwing2015,talAerobaticTrajectoryGeneration2023,airimitoaieConvertibleAircraftDynamic2018}.
For 2D Lift+Cruise vehicles, differential flatness introduces an ambiguity between the pitch angle and total thrusts:
For a given reference position command $p_{\text{ref}}$,
the reference pitch angle $\theta_{\text{ref}}$ should be determined to assign total thrusts $T_{r}$ and $T_{p}$~\cite{bullockReferenceCommandOptimization2024}.
In~\cite{bullockReferenceCommandOptimization2024},
an $\ell_{1}$-norm minimization problem was formulated to resolve this ambiguity.
The optimization problem can be written as follows:
\begin{equation}
\label{eq:opt}
\begin{split}
\min_{T_{r}(t), T_{p}(t), \delta_{e}(t), \theta_{\text{ref}}(t)} T_{r}(t) &+ T_{p}(t),
\\
\text{subject to} \quad T_{r}(t) &\geq 0 \quad \text{(nonnegative total vertical thrust)},
\\
T_{p}(t) &\geq 0 \quad \text{(nonnegative total horizontal thrust)},
\\
\theta_{\text{ref}}(t) &\in [\underline{\theta}, \overline{\theta}] \quad \text{(pitch constraint)},
\\
\delta_{e}(t) &\in [\underline{\delta}_{e}, \overline{\delta}_{e}] \quad \text{(elevator constraint)},
\\
\begin{bmatrix}
    T_{p} (t) \\ -T_{r} (t)
\end{bmatrix} &= \begin{bmatrix}
    F_{\text{des, x}} (t) + D_{1} (t) \\ F_{\text{des, y}} (t) + L (t) - D_{2} (t)
\end{bmatrix} \quad \text{(desired force constraint)}.
\end{split}
\end{equation}
Note that the cost function of the optimization problem \eqref{eq:opt}
can be regarded as the $\ell_{1}$-norm of total thrusts due to the nonnegativity constraints.
The $\ell_{1}$-norm minimization encourages the sparsity between vertical and horizontal total thrusts, $T_{r}$ and $T_{p}$, respectively,
implying transitions between hover and cruise flight modes.
The optimization problem in \eqref{eq:opt}
can be viewed as a constrained \textit{time-varying} optimization problem as described in section \ref{sec:tv_opt}.
That is, the cost function and constraints depend on time,
and the solution is also a function of time.
In~\cite{bullockReferenceCommandOptimization2024},
this time-varying optimization is approximately solved offline as follows:
the time interval of interest $[t_{0}, t_{f}]$ is divided into multiple time segments,
which can be specified by time stamps $\{t_{i} \}_{i \in \mathcal{I}}$.
Then, for each time stamp $t_{i}$,
the optimization problem in \eqref{eq:opt} is solved pointwise-in-time by setting $t=t_{i}$.
After solving the optimization problems for all time stamps $t_{i}$'s,
the reference commands and control inputs,
$\theta_{\text{ref}}(t)$, $T_{r}(t)$, $T_{p}(t)$, and $\delta_{e}(t)$ for all $t \in [t_{0}, t_{f}]$,
are interpolated over time.

In this study, the above method will be called open-loop optimization (OL-Opt) method.
We identified the limitations of OL-Opt as follows:
\begin{itemize}
    \item  (Non-smooth reference command)
    OL-Opt solves multiple optimization problems for each time stamp and interpolates the solutions over time. Therefore, the resulting solutions can drastically change and be non-smooth.
    \item (Offline reference command generation)
    OL-Opt generates reference commands offline. Closed-loop systems with tracking controllers may introduce accumulated tracking errors due to input saturation, and reference commands need to be adjusted online.
\end{itemize}
To overcome the above limitations,
closed-loop time-varying optimization (CL-TVOpt) method will be proposed in the next section.
The CL-TVOpt exhibits the following beneficial characteristics:
\begin{itemize}
    \item (Smooth reference command) CL-TVOpt updates reference commands continuously over time, which are solutions to an ODE. Therefore, CL-TVOpt can generate a smooth reference command.
    \item (Online reference command generation)
    CL-TVOpt can update the reference commands forward in time by solving an ODE, enabling online reference command generation.
\end{itemize}

\section{Closed-Loop Time-Varying Optimization for Smooth Reference Command Generation}
\label{sec:main}
In this section, the closed-loop time-varying optimization (CL-TVOpt) method is proposed for smooth reference command generation of the Lift+Cruise vehicle.

\subsection{Algorithm}
\begin{figure}
    \centering
    \includegraphics[width=0.45\linewidth]{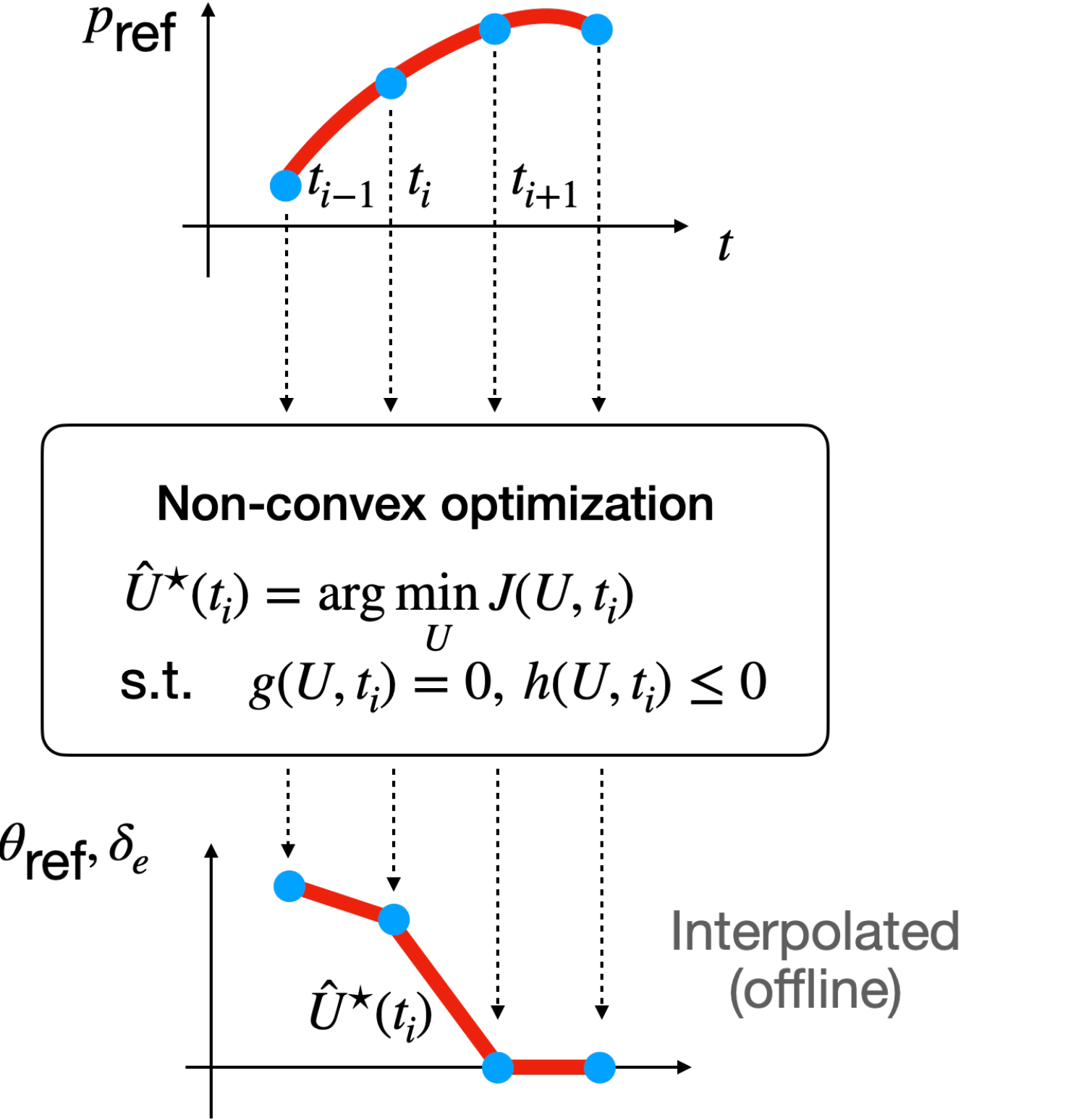}
    \includegraphics[width=0.45\linewidth]{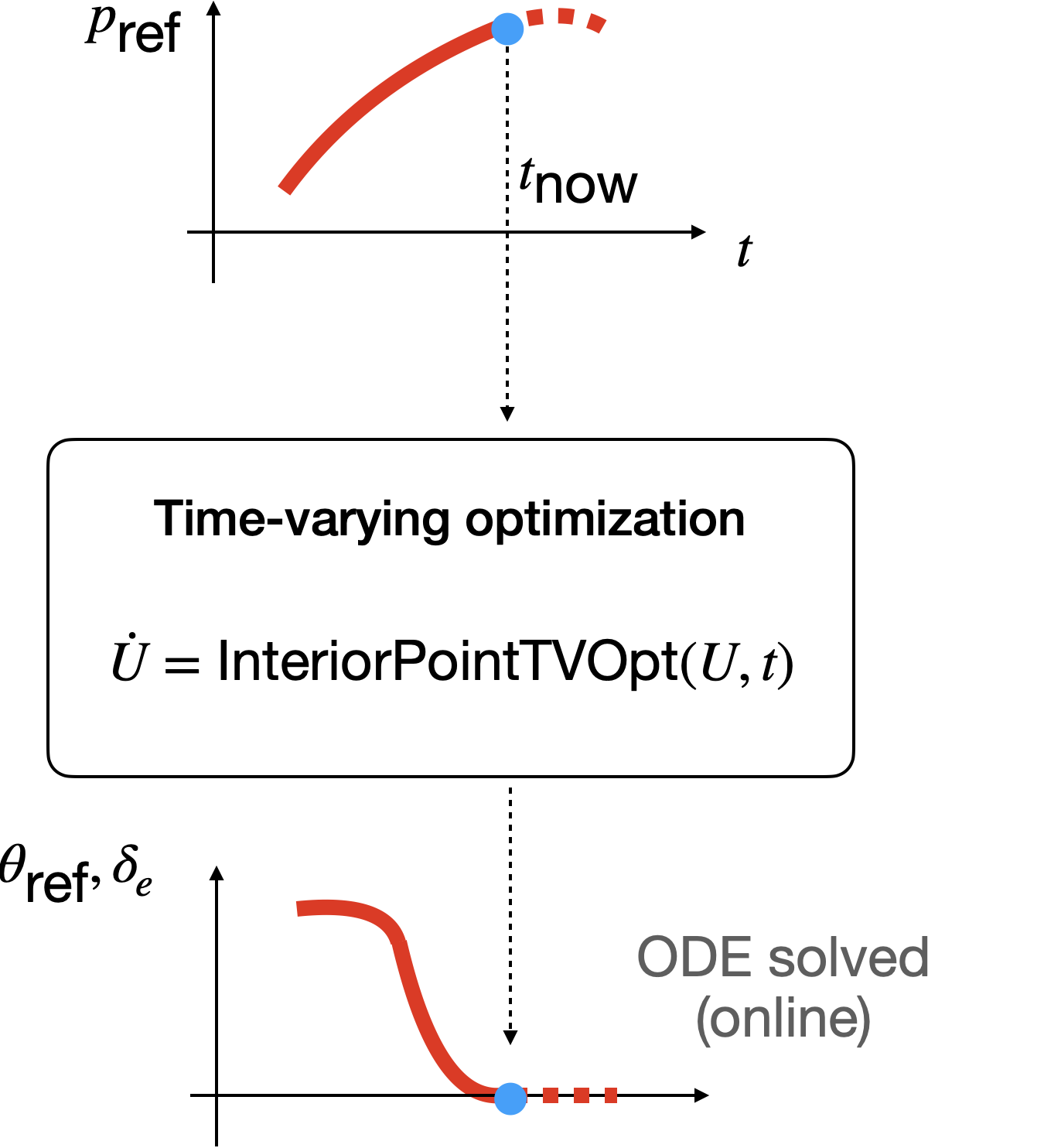}
    \caption{Conceptual differences between OL-Opt (left) and the proposed CL-TVOpt (right).}
    \label{fig:concept}
\end{figure}

First,
the right-hand side of the equality constraint in \eqref{eq:opt} is substituted into the cost function and inequality constraints to replace the total thrusts $T_{r}(t)$ and $T_{p}(t)$.
Rearranging equations yield
\begin{equation}
\label{eq:tvopt1}
\begin{split}
\min_{U(t) := [\delta_{e}(t), \theta_{\text{ref}}(t)]^{\intercal}} J(U(t), t) := F_{\text{des}, x}(t) &- F_{\text{des, y}}(t) + D_{1}(t) - L(t) + D_{2}(t)
\\
\text{subject to} \quad h_{1}(U(t), t) &:= F_{\text{des, y}}(t) + L(t) - D_{2}(t) \leq 0,
\\
h_{2}(U(t), t) &:= -F_{\text{des, x}}(t) - D_{1}(t) \leq 0,
\\
h_{3}(U(t), t) &:= \underline{\theta} - \theta_{\text{ref}}(t) \leq 0,
\\
h_{4}(U(t), t) &:= \theta_{\text{ref}}(t) - \overline{\theta} \leq 0,
\\
h_{5}(U(t), t) &:= \underline{\delta}_{e} - \delta_{e}(t) \leq 0,
\\
h_{6}(U(t), t) &:= \delta_{e}(t) -\overline{\delta}_{e} \leq 0.
\end{split}
\end{equation}
Consider the following interior-point formulation:
\begin{equation}
\label{eq:tvopt2}
\begin{split}
\min_{U(t)} \Phi(U(t), t) := J(U(t), t) - \sum_{i=1}^{6} \frac{1}{c_{i}} \log \left( \epsilon_{i} -h_{i}(U(t), t) \right),
\end{split}
\end{equation}
where $\epsilon_{i} > 0$ are small positive constants to avoid numerical instability for logarithmic functions around zero,
and $c_{i} > 0$ are positive constants for approximating the original optimization problem with log-barrier functions.
In the interior-point method, the logarithmic functions prevent the violation of inequality constraints.
The proposed method, CL-TVOpt, utilizes the prediction-correction method~\cite{fazlyabPredictionCorrectionInteriorPointMethod2018} as follows:
\begin{equation}
    \label{eq:update_law}
    \dot{U}(t) = - \nabla_{UU}\Phi^{-1} (U(t), t) \left( P \nabla_{U}\Phi(U(t), t) + \nabla_{U t}\Phi(U(t), t)\right ),
\end{equation}
where $P \succeq \alpha I$ determines the convergence rate,
and the partial derivatives of $\nabla_{U}\Phi(U, t)$ with respect to $U$ and $t$ are denoted by $\nabla_{UU}\Phi(U, t)$ and $\nabla_{U t} \Phi(U, t)$, respectively.
Although the original paper shows the convergence for time-varying convex optimization~\cite{fazlyabPredictionCorrectionInteriorPointMethod2018},
it can readily be shown under the update law in \eqref{eq:update_law}
that the gradient converges to zero exponentially with the rate of $\alpha > 0$, i.e.,
$\lVert \nabla_{U}\Phi(U(t), t) \rVert \leq \exp(- \alpha t) \lVert \nabla_{U}\Phi(U(0), 0) \rVert $.
This implies that the optimization variables converge to a time-varying local minimizer.
The conceptual comparison between OL-Opt and the proposed method is illustrated in \autoref{fig:concept}.

\begin{remark}[Online reference command generation]
In OL-Opt,
the optimization problems for each time stamp are solved in advance,
and then the reference command is generated by interpolating the solutions.
This implies that the OL-Opt may not be used to generate a reference command online.
Unlike OL-Opt,
the proposed CL-TVOpt can generate a reference command online by forming a closed-loop system with the update law characterized as an ODE in \eqref{eq:update_law}.
\end{remark}

\section{Numerical Simulation}
\label{sec:sim}
\begin{figure}
    \centering
    \includegraphics[width=0.65\linewidth]{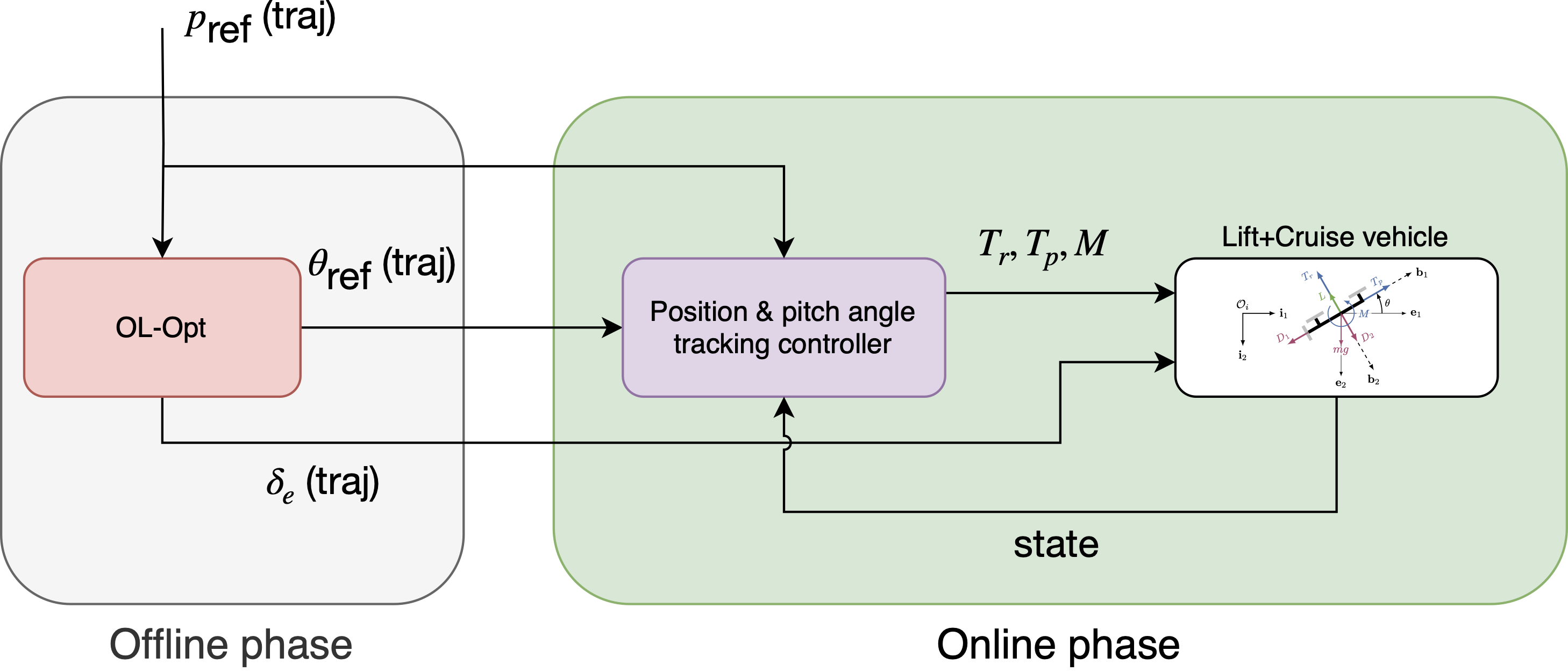}
    \\
    (a) OL-Opt~\cite{bullockReferenceCommandOptimization2024}
    \\
    \includegraphics[width=0.65\linewidth]{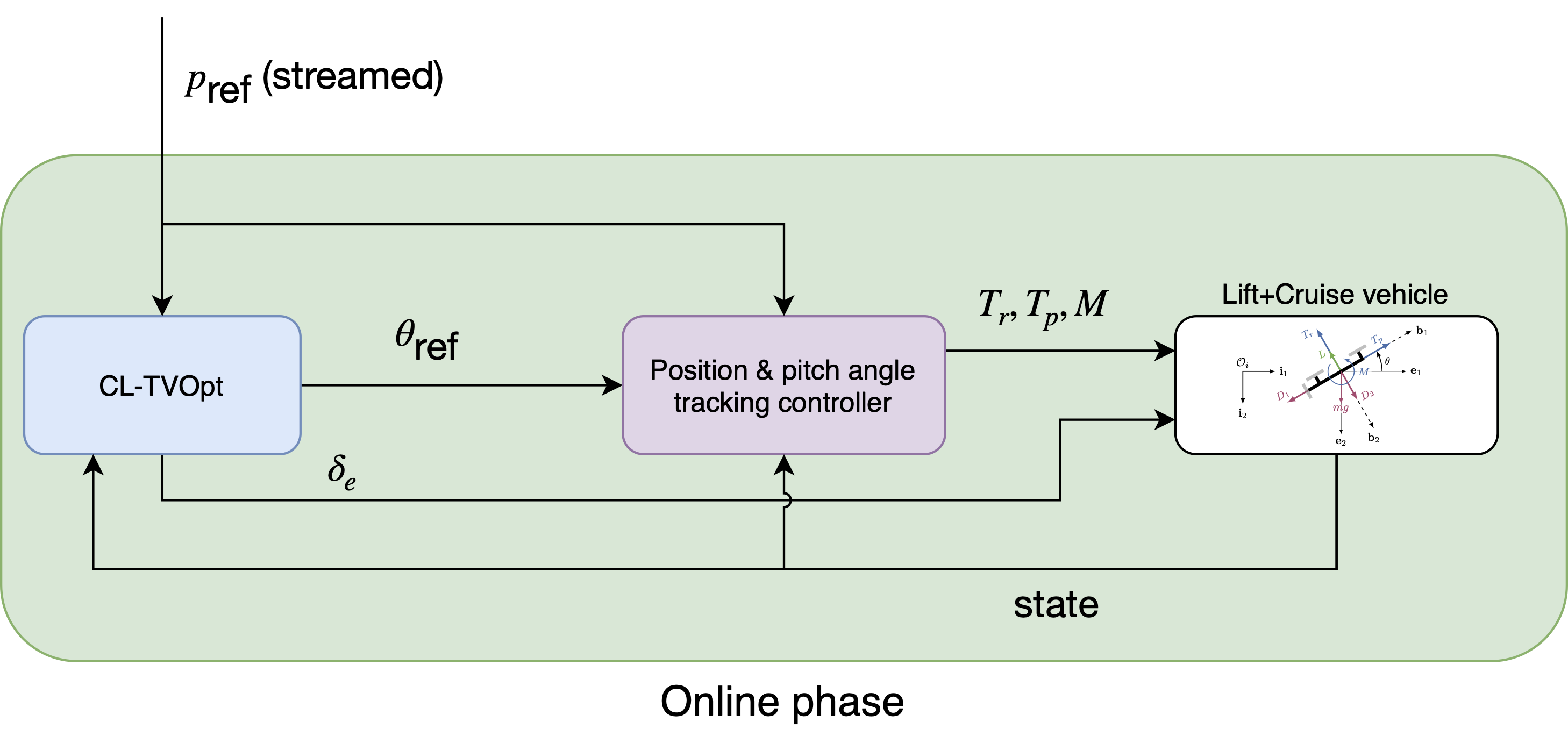}
    \\
    (b) CL-TVOpt (proposed)
    \\
    \caption{Block diagrams of each method in simulation}
    \label{fig:how_to_sim}
\end{figure}

In this section, the proposed CL-TVOpt will be compared to OL-Opt in two scenarios: i) Hover-to-Cruise and ii) Cruise-to-Hover.
The time horizon of each scenario is given by $t_{0} = 0$ s and $t_{f} = 125$ s.
For OL-Opt, the time stamps are made evenly with the time step of $0.1$ s.
Then, the elevator deflection angle $\delta_{e}$ and reference pitch angle command $\theta_{\text{ref}}$ are interpolated over time.
For both OL-Opt and the proposed CL-TVOpt, the position and pitch angle tracking controllers are implemented for stability.
The parameters for the Lift+Cruise model are borrowed from~\cite{bullockReferenceCommandOptimization2024}.
A B\'ezier curve is used for the reference position command $p_{\text{ref}}(t) = [x_{\text{ref}}(t), z_{\text{ref}}(t)]^{\intercal}$.
The initial condition is given as $p(0)=\dot{p}(0)=\theta(0)=\dot{\theta}(0)=0$.
The input saturation is considered in numerical simulation with bounds as $\delta_{e}(t) \in [\underline{\delta}_{e}, \overline{\delta}_{e}] = [-30, 30] \text{ (deg)}$,
$T_{r} \geq 0$,
and $T_{p} \geq 0$.
\autoref{fig:how_to_sim} illustrates the block diagrams of each method for numerical simulation.
The code is written in Julia~\cite{bezansonJuliaFreshApproach2017}.
Other details of numerical simulation can be found in the Appendix.

\subsection{Scenario 1: Hover-to-Cruise}
\label{sec:scenario1}
\begin{figure}[t!]
    \centering
    \includegraphics[width=0.95\linewidth]{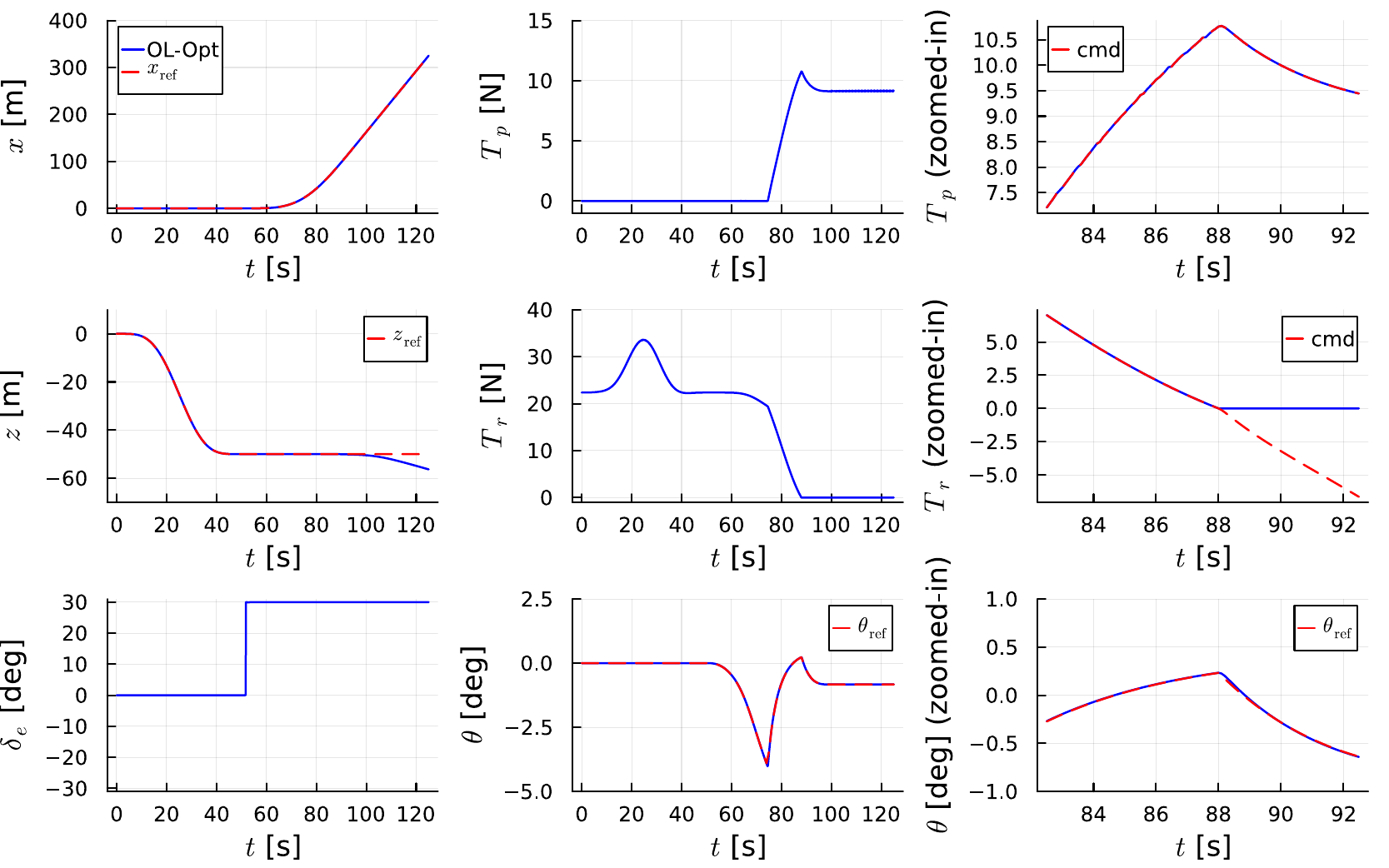}
    \caption{
    Scenario 1 (Hover-to-Cruise):
    The result of OL-Opt.
    }
    \label{fig:h2c_ol_opt_linear}
\end{figure}

\begin{figure}[t!]
    \centering
    \includegraphics[width=0.95\linewidth]{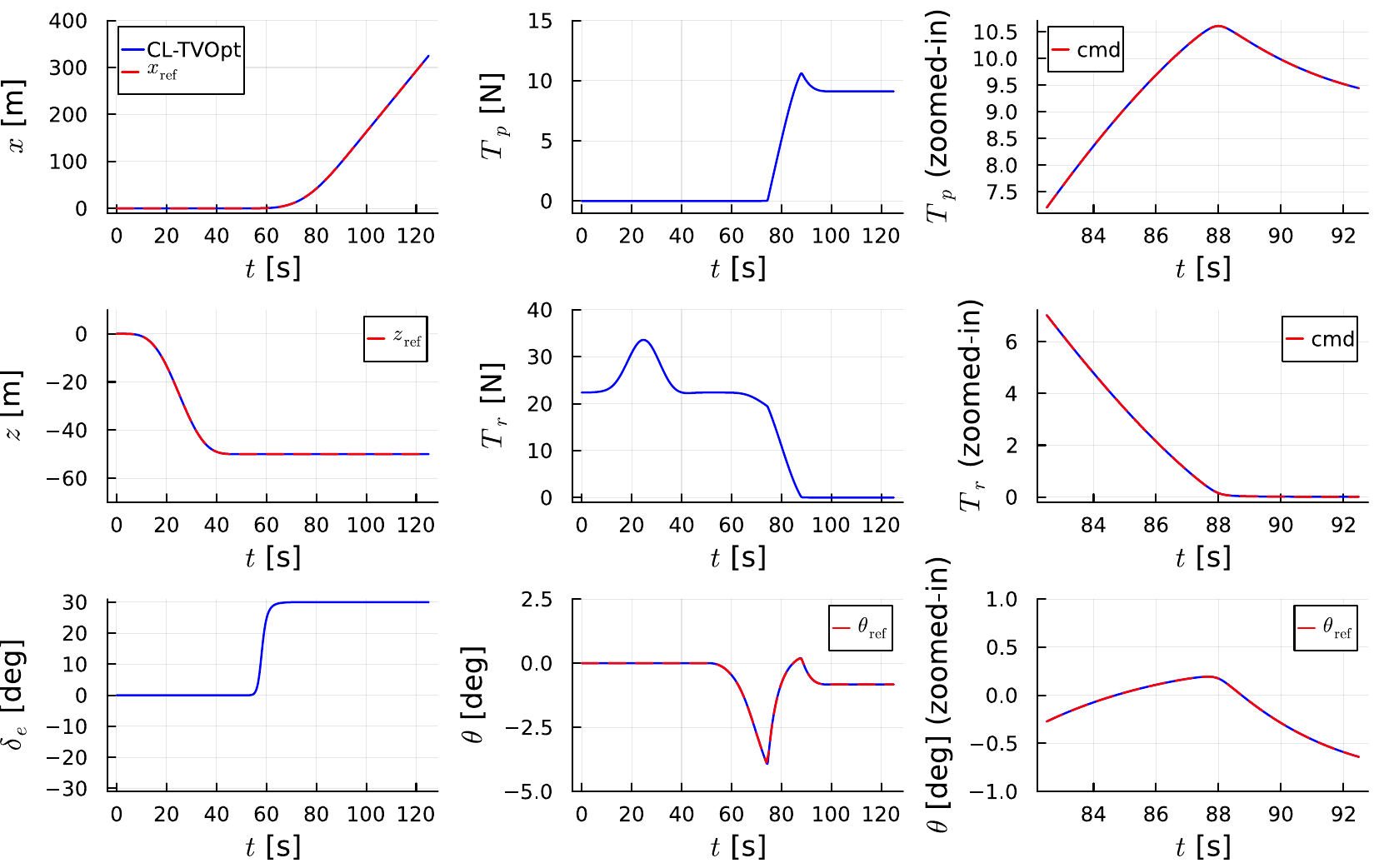}
    \caption{
    Scenario 1 (Hover-to-Cruise):
    The result of CL-TVOpt (proposed).
    }
    \label{fig:h2c_cl_tvopt}
\end{figure}

In this section, the Hover-to-Cruise scenario is considered.
\autoref{fig:h2c_ol_opt_linear} and \autoref{fig:h2c_cl_tvopt} show the simulation results of OL-Opt and the proposed CL-TVOpt, respectively.
In \autoref{fig:h2c_ol_opt_linear},
the total thrusts show sparsity as expected from $\ell_{1}$-norm minimization.
However,
the OL-Opt shows a non-smooth transition of $T_{r}$ from positive to zero at about $t=90$ s.
This non-smoothness gives a slight overshoot in $z$.
Because $T_{r}$ is already saturated at zero, the tracking controller cannot compensate for the slight error, which increases the $z$ error and $T_{r}$ command gradually.
Also, the elevator shows a bang-bang control behavior at about $t=50$ s.
On the other hand,
as shown in \autoref{fig:h2c_cl_tvopt},
the proposed CL-TVOpt method provides smooth reference commands for pitch angle and elevator deflection angle with sparsity in total thrusts, achieving smooth transitions from hover to cruise.

\subsection{Scenario 2: Cruise-to-Hover}
\label{sec:scenario2}
\begin{figure}[t!]
    \centering
    \includegraphics[width=0.95\linewidth]{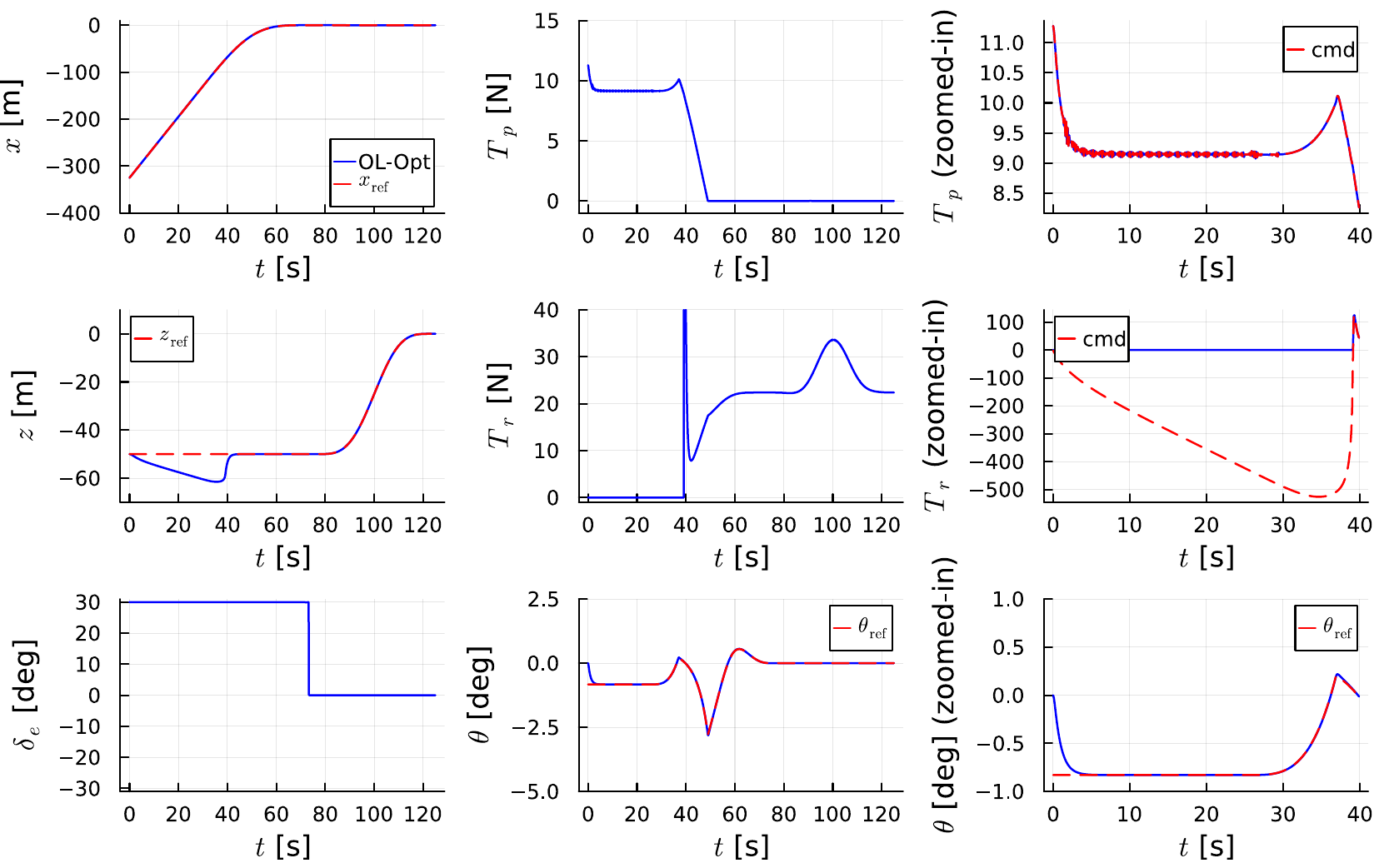}
    \caption{
    Scenario 2 (Cruise-to-Hover):
    The result of OL-Opt.
    }
    \label{fig:c2h_ol_opt_linear}
\end{figure}

\begin{figure}[t!]
    \centering
    \includegraphics[width=0.95\linewidth]{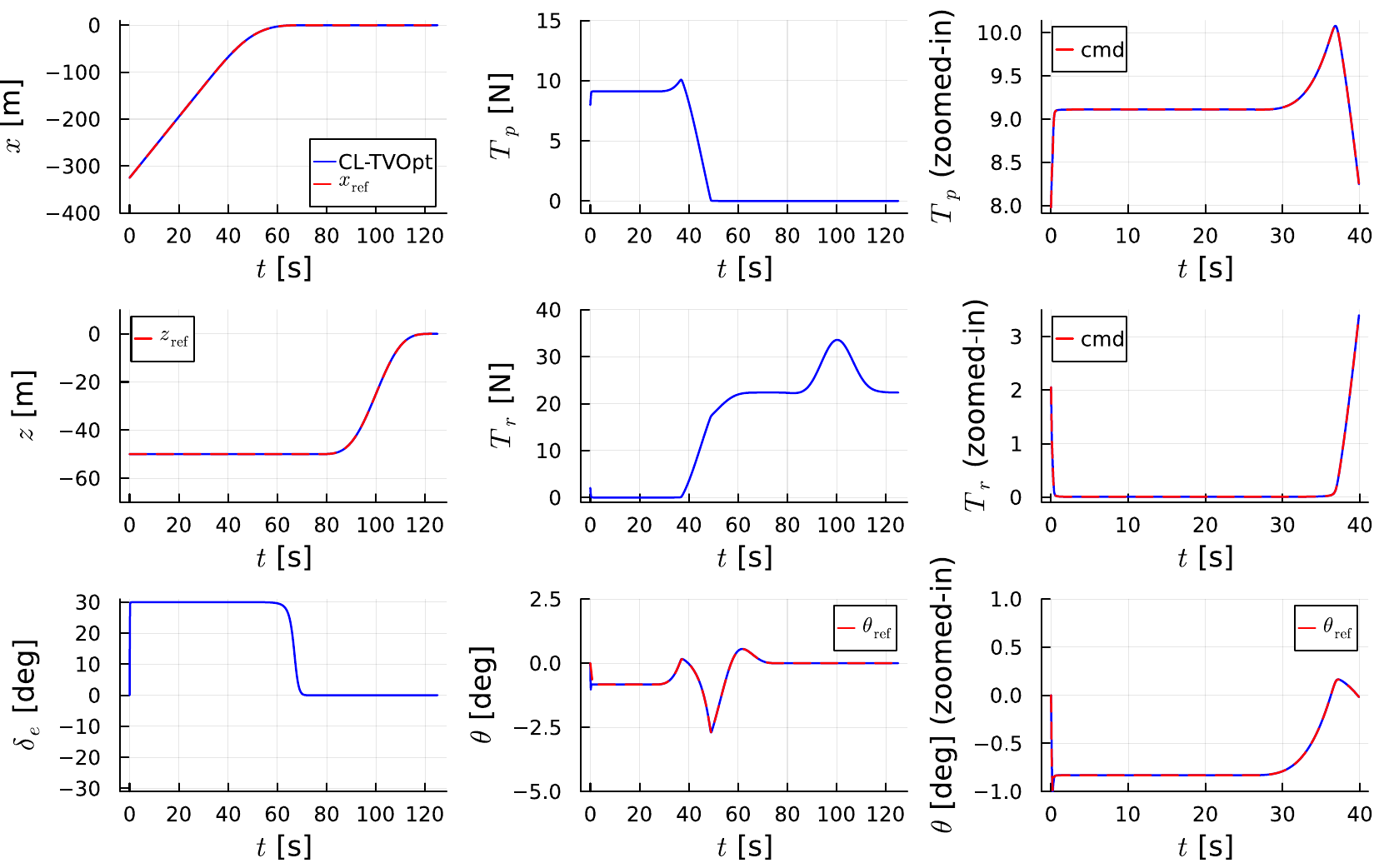}
    \caption{
    Scenario 2 (Cruise-to-Hover):
    The result of CL-TVOpt (proposed).
    }
    \label{fig:c2h_cl_tvopt}
\end{figure}

In this section, the Cruise-to-Hover scenario is considered.
\autoref{fig:c2h_ol_opt_linear} and \autoref{fig:c2h_cl_tvopt} show the simulation results of OL-Opt and the proposed CL-TVOpt,
respectively.
From the Cruise-to-Hover scenario shown in \autoref{fig:c2h_ol_opt_linear},
OL-Opt initially increases the height ($-z$) due to the initial error in pitch angle $\theta$.
Such initial tracking error is typically inevitable for offline reference command generation methods.
However, the total thrust of the vertical rotors, $T_{r}$, is already saturated at zero.
Similar to the Hover-to-Cruise scenario,
the Lift+Cruise vehicle cannot compensate for the $z$ error due to the saturation of $T_{r}$.
This gradually increases the $T_{r}$ command by the tracking controller,
yielding an impulse-like control input in $T_{r}$ at about $t=40$ s through the transition,
which can be harmful for safe and reliable transition in practice.
Also, the elevator deflection angle changes abruptly at about $t=70$ s.
On the contrary,
as shown in \autoref{fig:c2h_cl_tvopt},
the proposed CL-TVOpt method can avoid the initial tracking error by generating smooth reference commands online.
The proposed method also shows the sparsity of total thrusts in this scenario.

\autoref{tab:elapsed_time} shows the mean elapsed times for each method, averaged over ten runs.
For each run, the mean elapsed time is calculated by dividing the total elapsed time by the number of points.
The number of points is $(t_{f} - t_{0}) / \Delta t$ where $\Delta t$ is the time step for optimization.
For the OL-Opt, only the pre-computation of optimization for multiple time stamps is considered for the total elapsed time with $\Delta t = 0.1$ s, i.e., $N= 1\text{,}250$.
For the proposed CL-TVOpt, the time to solve the entire ODE (including state propagation) is considered for the total elapsed time, which overestimates the required computation for the CL-TVOpt.
Note that the $\Delta t$ of the CL-TVOpt is the same as the simulation time step of $0.01$ s, i.e., $N=12\text{,}250$.
For Hover-to-Cruise scenario, the overestimate of the proposed method's mean elapsed time is smaller ($12$\%) than the mean elapsed time of the OL-Opt.
For Cruise-to-Hover scenario, the overestimate of the proposed method's mean elapsed time is also smaller ($13$\%) than the mean elapsed time of the OL-Opt.
This is because OL-Opt requires iterations to converge for solving optimization problems,
while the proposed CL-TVOpt replaces solving optimization problems by solving the ordinary differential equation in \eqref{eq:update_law}.
\begin{table}[!h]
    \caption{Mean elapsed time}
    \centering
    \begin{tabular}{c|c|c}
    \hline
    & Hover-to-Cruise & Cruise-to-Hover \\
    \hline
    \hline
    OL-Opt$^{*}$ & $3.246$ ms & $2.826$ ms
    \\
    \hline
    CL-TVOpt$^{**}$ (proposed) & $< \mathbf{0.387}$ ms & $< \mathbf{0.377}$ ms
    \\
    \hline
    \multicolumn{3}{l}{%
    \begin{minipage}{10cm}%
        $^{*}$: Elapsed time only for pre-computation of optimization (interpolation not included; section \ref{sec:ol_opt})
        \\
        $^{**}$: Elapsed time for the ODE solver including state propagation
    \end{minipage}%
        }\\
    \end{tabular}
    \label{tab:elapsed_time}
\end{table}
In short,
compared to the existing open-loop method,
the proposed method can generate smooth reference pitch commands and control inputs online, enabling smooth transitions of Lift+Cruise aircraft between hover and cruise flight modes.

\section{Conclusion}
\label{sec:conclusion}
In this paper,
a time-varying optimization-based online reference command generation method was proposed for the transition flight of Lift+Cruise type vertical take-off and landing (VTOL) aircraft.
For the transition flight,
a time-varying $\ell_{1}$-norm minimization formulation was utilized to encourage the sparsity in vertical and horizontal total thrusts.
The proposed method utilizes the prediction-correction interior-point time-varying optimization method.
The resulting algorithm continuously updates the optimization variables forward in time by solving an ordinary differential equation (ODE),
enabling online smooth reference command generation.
Simulation results showed that the proposed method can generate smooth reference pitch command and control inputs.

Future works include extensions to
three-dimensional Lift+Cruise vehicles with realistic aerodynamic effects
and the consideration of uncertainty.

\section*{Appendix}
\subsection{Simulation details}
All ODEs are solved by \texttt{Tsit5} (Tsitouras 5/4 Runge-Kutta method) with a fixed time step of $0.01$ s using DifferentialEquations.jl~\cite{rackauckas2017differentialequations}.
For OL-Opt, Ipopt~\cite{wachter2006implementation} is used for non-convex optimization,
and scaled B-spline linear interpolation is used by Interpolations.jl~\cite{interpolations_jl}.
The parameters $c_{i}$ and $\epsilon_{i}$ in \eqref{eq:tvopt2} are set as $c_{i}=10^{3}$ and $\epsilon_{i} = 10^{-3}$ for all $i \in \{ 1 \ldots, 6\}$.
The gain matrix $P$ in \eqref{eq:update_law} is set as the identity matrix, i.e., $P = I_{2}$.

In numerical simulation,
the following position tracking controller is used:
\begin{equation}
\label{eq:position_controller}
\begin{bmatrix}
    T_{r}(t) \\ -T_{p}(t)
\end{bmatrix}
=
    \begin{bmatrix}
    D_{1}(t) \\ L(t)-D_{2}(t)
    \end{bmatrix}
    + m R(\theta(t)) \left(
        -g e_{2}
        + \ddot{p}_{\text{ref}}(t)
        - k_{D, p} (\dot{p}(t) - \dot{p}_{\text{ref}}(t))
        - k_{P, p} (p(t) - p_{\text{ref}}(t))
    \right).
\end{equation}
Under the position tracking controller,
the resulting error dynamics can be written as
\begin{equation}
    \ddot{\tilde{p}}(t) + k_{D, p} \dot{\tilde{p}}(t) + k_{P, p} \tilde{p}(t) = 0,
\end{equation}
where $\tilde{p}(t) := p(t) - p_{\text{ref}}(t) $ is the position error.
The gains of the position tracking controller are set as
$k_{D, p} = 20$ and $k_{P, p} = 20$.

For pitch angle,
the following tracking controller is used:
\begin{equation}
    M = J \left(
        \ddot{\theta}_{\text{ref}} - k_{P, \theta} (\dot{\theta} - \dot{\theta}_{\text{ref}}) - k_{D, \theta} (\theta - \theta_{\text{ref}})
    \right),
\end{equation}
which results in the following error dynamics:
    \begin{equation}
    \ddot{\tilde{\theta}}(t) + k_{D, \theta} \dot{\tilde{\theta}}(t) + k_{P, \theta} \tilde{\theta}(t) = 0,
\end{equation}
where $\tilde{\theta}(t) := \theta(t) - \theta_{\text{ref}}(t) $ is the pitch angle error.
The gains are set as $k_{P, \theta} = 10$ and $k_{D, \theta} =10$ in simulation.
For the OL-Opt,
the derivatives of reference pitch command, $\dot{\theta}_{\text{ref}}$ and $\ddot{\theta}_{\text{ref}}$ are obtained by numerical differentiation using ForwardDiff.jl~\cite{RevelsLubinPapamarkou2016}.
For the proposed CL-TVOpt,
$\dot{\theta}_{\text{ref}}$ can be obtained from \eqref{eq:update_law},
and $\ddot{\theta}_{\text{ref}}$ is approximated by adding a filter,
that is,
$\ddot{\theta}_{\text{ref}} \approx \frac{\hat{\dot{\theta}}_{\text{ref}}(t) - \dot{\theta}_{\text{ref}}(t) }{\tau}$ with the filter state $\hat{\dot{\theta}}_{\text{ref}}$ governed by the following equation:
\begin{equation}
    \frac{d}{dt} \hat{\dot{\theta}}(t) = \frac{ \dot{\theta}_{\text{ref}}(t) - \hat{\dot{\theta}}_{\text{ref}}(t) }{\tau},
\end{equation}
where $\tau > 0$ is the time constant.
The time constant is set as $\tau = 0.01$ in simulation.

\section*{Acknowledgments}
This work is supported by NASA under the Cooperative Agreement 80NSSC20M0229 and University Leadership Initiative grant 80NSSC22M0070.

\bibliography{main}

\begin{thebibliography}{17}
\newcommand{\enquote}[1]{``#1''}
\providecommand{\natexlab}[1]{#1}
\providecommand{\url}[1]{\texttt{#1}}
\providecommand{\urlprefix}{URL }
\expandafter\ifx\csname urlstyle\endcsname\relax
  \providecommand{\doi}[1]{\discretionary{}{}{}https://doi.org/#1}\else
  \providecommand{\doi}[1]{\discretionary{}{}{}\urlstyle{rm}\url{https://doi.org/#1}}\fi

\bibitem[{Simmons et~al.(2021)Simmons, Buning, and
  Murphy}]{simmonsFullEnvelopeAeroPropulsiveModel2021}
Simmons, B.~M., Buning, P.~G., and Murphy, P.~C., \enquote{Full-{Envelope}
  {Aero}-{Propulsive} {Model} {Identification} for {Lift}+{Cruise} {Aircraft}
  {Using} {Computational} {Experiments},} \emph{{AIAA} {AVIATION} 2021
  {FORUM}}, American Institute of Aeronautics and Astronautics, VIRTUAL EVENT,
  2021.
\newblock \doi{10.2514/6.2021-3170}.

\bibitem[{Bullock et~al.(2024)Bullock, Cheng, Patterson, Acheson, Hovakimyan,
  and Gregory}]{bullockReferenceCommandOptimization2024}
Bullock, J.~L., Cheng, S., Patterson, A., Acheson, M.~J., Hovakimyan, N., and
  Gregory, I.~M., \enquote{Reference {Command} {Optimization} for the
  {Transition} {Flight} {Mode} of a {Lift} {Plus} {Cruise} {Vehicle},}
  \emph{{AIAA} {SCITECH} 2024 {Forum}}, American Institute of Aeronautics and
  Astronautics, Orlando, FL, 2024.
\newblock \doi{10.2514/6.2024-0721}.

\bibitem[{Lee and Kim(2024)}]{leeMPCBasedLongitudinalControl2024}
Lee, J., and Kim, Y., \enquote{{MPC}-{Based} {Longitudinal} {Control} {Design}
  for a {Fixed}-{Wing} {VTOL} {UAV} in {Transition} {Flight},} \emph{{AIAA}
  {SCITECH} 2024 {Forum}}, American Institute of Aeronautics and Astronautics,
  Orlando, FL, 2024.
\newblock \doi{10.2514/6.2024-0141}.

\bibitem[{Kim et~al.(2023)Kim, Lee, Ko, and Kim}]{kimVTOLAircraftOptimal2023}
Kim, J., Lee, H., Ko, D., and Kim, Y., \enquote{{VTOL} {Aircraft} {Optimal}
  {Gain} {Prediction} via {Parameterized} {Log}-{Sum}-{Exp} {Networks},}
  \emph{European {Control} {Conference} ({ECC})}, Bucharest, Romania, 2023.
\newblock \doi{10.23919/ECC57647.2023.10178276}.

\bibitem[{Acheson(2024)}]{guam_v11}
Acheson, M.~J., \emph{Generic-Urban-Air-Mobility-GUAM}, NASA Langley Research
  Center, Hampton, VA, v1.1 ed., May 2024.
\newblock
  \urlprefix\url{https://github.com/nasa/Generic-Urban-Air-Mobility-GUAM}.

\bibitem[{Sun et~al.(2022)Sun, Romero, Foehn, Kaufmann, and
  Scaramuzza}]{sunComparativeStudyNonlinear2022}
Sun, S., Romero, A., Foehn, P., Kaufmann, E., and Scaramuzza, D., \enquote{A
  {Comparative} {Study} of {Nonlinear} {MPC} and
  {Differential}-{Flatness}-{Based} {Control} for {Quadrotor} {Agile}
  {Flight},} \emph{IEEE Transactions on Robotics}, Vol.~38, No.~6, 2022, pp.
  3357--3373.
\newblock \doi{10.1109/TRO.2022.3177279}.

\bibitem[{Fazlyab et~al.(2018)Fazlyab, Paternain, Preciado, and
  Ribeiro}]{fazlyabPredictionCorrectionInteriorPointMethod2018}
Fazlyab, M., Paternain, S., Preciado, V.~M., and Ribeiro, A.,
  \enquote{Prediction-{Correction} {Interior}-{Point} {Method} for
  {Time}-{Varying} {Convex} {Optimization},} \emph{IEEE Transactions on
  Automatic Control}, Vol.~63, No.~7, 2018, pp. 1973--1986.
\newblock \doi{10.1109/TAC.2017.2760256}.

\bibitem[{Faessler et~al.(2017)Faessler, Falanga, and
  Scaramuzza}]{faesslerThrustMixingSaturation2017}
Faessler, M., Falanga, D., and Scaramuzza, D., \enquote{Thrust {Mixing},
  {Saturation}, and {Body}-{Rate} {Control} for {Accurate} {Aggressive}
  {Quadrotor} {Flight},} \emph{IEEE Robotics and Automation Letters}, Vol.~2,
  No.~2, 2017, pp. 476--482.
\newblock \doi{10.1109/LRA.2016.2640362}.

\bibitem[{Mellinger and Kumar(2011)}]{mellingerMinimumSnapTrajectory2011a}
Mellinger, D., and Kumar, V., \enquote{Minimum Snap Trajectory Generation and
  Control for Quadrotors,} \emph{2011 {IEEE} {International} {Conference} on
  {Robotics} and {Automation}}, IEEE, Shanghai, China, 2011, pp. 2520--2525.
\newblock \doi{10.1109/ICRA.2011.5980409}.

\bibitem[{Bry et~al.(2015)Bry, Richter, Bachrach, and
  Roy}]{bryAggressiveFlightFixedwing2015}
Bry, A., Richter, C., Bachrach, A., and Roy, N., \enquote{Aggressive Flight of
  Fixed-wing and Quadrotor Aircraft in Dense Indoor Environments,} \emph{The
  International Journal of Robotics Research}, Vol.~34, No.~7, 2015, pp.
  969--1002.
\newblock \doi{10.1177/0278364914558129}.

\bibitem[{Tal et~al.(2023)Tal, Ryou, and
  Karaman}]{talAerobaticTrajectoryGeneration2023}
Tal, E., Ryou, G., and Karaman, S., \enquote{Aerobatic {Trajectory}
  {Generation} for a {VTOL} {Fixed}-{Wing} {Aircraft} {Using} {Differential}
  {Flatness},} \emph{IEEE Transactions on Robotics}, Vol.~39, No.~6, 2023, pp.
  4805--4819.
\newblock \doi{10.1109/TRO.2023.3301312}.

\bibitem[{Airimitoaie et~al.(2018)Airimitoaie, Aguilar, Lavigne, Farges, and
  Cazaurang}]{airimitoaieConvertibleAircraftDynamic2018}
Airimitoaie, T.-B., Aguilar, G.~P., Lavigne, L., Farges, C., and Cazaurang, F.,
  \enquote{Convertible Aircraft Dynamic Modelling and Flatness Analysis,}
  \emph{IFAC-PapersOnLine}, Vol.~51, No.~2, 2018, pp. 25--30.
\newblock \doi{10.1016/j.ifacol.2018.03.005}.

\bibitem[{Bezanson et~al.(2017)Bezanson, Edelman, Karpinski, and
  Shah}]{bezansonJuliaFreshApproach2017}
Bezanson, J., Edelman, A., Karpinski, S., and Shah, V.~B., \enquote{Julia: {A}
  {Fresh} {Approach} to {Numerical} {Computing},} \emph{SIAM Review}, Vol.~59,
  No.~1, 2017, pp. 65--98.
\newblock \doi{10.1137/141000671}.

\bibitem[{Rackauckas and Nie(2017)}]{rackauckas2017differentialequations}
Rackauckas, C., and Nie, Q., \enquote{Differentialequations.jl--A Performant
  and Feature-rich Ecosystem for Solving Differential Equations in Julia,}
  \emph{Journal of Open Research Software}, Vol.~5, No.~1, 2017, p.~15.

\bibitem[{W{\"a}chter and Biegler(2006)}]{wachter2006implementation}
W{\"a}chter, A., and Biegler, L.~T., \enquote{On the Implementation of an
  Interior-point Filter Line-search Algorithm for Large-scale Nonlinear
  Programming,} \emph{Mathematical programming}, Vol. 106, 2006, pp. 25--57.

\bibitem[{Kittisopikul and Holy(2022)}]{interpolations_jl}
Kittisopikul, M., and Holy, T.~E., \enquote{JuliaMath/Interpolations.jl,} ,
  2022.
\newblock \urlprefix\url{https://github.com/JuliaMath/Interpolations.jl}.

\bibitem[{{Revels} et~al.(2016){Revels}, {Lubin}, and
  {Papamarkou}}]{RevelsLubinPapamarkou2016}
{Revels}, J., {Lubin}, M., and {Papamarkou}, T., \enquote{Forward-Mode
  Automatic Differentiation in {J}ulia,} \emph{arXiv:1607.07892 [cs.MS]}, 2016.

\end{thebibliography}

\end{document}